\documentstyle{l-aa} 

\begin{document}

\title {Rapid X-ray variability of the superluminal source GRS1915+105. }

\author{ B. Paul\inst{1}, P. C. Agrawal\inst{1}, A. R. Rao\inst{1}, M. N. Vahia\inst{1}, J. S. Yadav \inst{1},
T. M. K. Marar\inst{2}, S. Seetha\inst{2} \and K. Kasturirangan \inst{2}}

\institute {Tata Institute of Fundamental Research, Homi Bhabha Road, Mumbai 400 005, India
\and ISRO Satellite Centre, Airport Road, Vimanpura P.O. Bangalore  560 017, India.} 

\offprints { B. Paul, {\it bpaul@tifrvax.tifr.res.in}}

\thesaurus{02.01.2; 02.02.1; 13.25.5; 08.09.2 GRS 1915+105}

\maketitle
\markboth{Paul et al.: Rapid X-ray variability of GRS 1915+105}{ }

\begin{abstract}
The superluminal X-ray transient source GRS 1915+105 was observed
during July 20-29, 1996 with the Indian  X-ray  Astronomy  Experiment
(IXAE)  on  the   Indian satellite IRS-P3 launched on March 21, 1996
from Shriharikota Range in  India. During our observations covering
the energy band 2-18 keV, we have
seen strong erratic  intensity variations  on  time  scale of 0.1s - 10s.
Quasi  Periodic Oscillations  (QPOs)  in a frequency range of
0.62 to 0.82  Hz  were detected with a rms fraction of about $9\%$.
The rapid X-ray intensity  variations 
in  GRS1915+105  are similar to those observed in  some other black  hole 
binaries and thus provide further support for the  hypothesis 
that this source is likely to be a black hole.  We discuss
the possible emission region and mechanism of the observed
quasi-periodic oscillations. Comparing the observed QPOs with the
ones observed in other neutron star and black-hole systems, we argue
that GRS1915+105 is possibly a black-hole.

\end{abstract}

\section{Introduction}

     The  Transient X-ray source GRS1915+105 was  discovered  with 
the  WATCH instrument onboard the GRANAT Observatory in 1992 (\cite{cast:94}).  Hard 
X-ray  studies  in  20-100 keV band have  shown  erratic  intensity 
variations   on  time  scales  of  days  and  months (\cite{fost:96}; \cite{sazo:94}). The source has been 
identified  with  a  superluminal  radio  source (\cite{rodr:93})  which   undergoes 
frequent flaring on a variety of time scales. Based on its  peculiar 
radio characteristics and superluminal motion, it  has  been 
termed as a micro-quasar (\cite{mira:94}). An infrared source with a jet at the same 
position  angle  as  seen in the radio counterpart, has  also  been 
identified  with this object (\cite{sams:96}). Based on its X-ray luminosity
which greatly  exceeds  the  Eddington limit and  radio  characteristics 
which  are similar to those of the radio loud quasars, it  has  been 
suggested  to be a black hole.  We have observed this  source  with 
the  Indian  X-ray  Astronomy  Experiment  (IXAE) and have detected
erratic  intensity variations  on  time  scale of 0.1s - 10s.  Strong
Quasi  Periodic Oscillations  (QPOs)  in a frequency range of 0.62 to
0.82  Hz  were also  detected unambiguously (\cite{agra:96b}).  In this letter we discuss
the possible emission region and mechanism of the observed
quasi-periodic oscillations. The strong and erratic intensity variations
are identical to the variations seen in other black-hole candidate X-ray
sources. The QPOs seen in GRS 1915+105 are compared with the same in other black
hole candidates and neutron star binaries.

\section{The IXAE and Observations}

     The  X-ray  observations  were made with  the  IXAE  which was launched  
onboard  the Indian Remote Sensing satellite-P3  (IRS-P3)
using  the Polar Satellite Launch Vehicle (PSLV) on March 21,  1996 
from  Shriharikota Range in India. The satellite is in  a 
circular orbit at an altitude of 830 km and inclination of
98$^\circ$. The  IXAE  includes three collimated Pointed Proportional
Counters  (PPCs) with an effective area of about 1200 cm$^2$ filled
with P-10 gas at  a 
pressure  of  800 torr.
Each PPC is a multi-anode, multi-layer detector with 54 anode cells 
of  size  11  mm  x 11 mm arranged in 3  layers  with  a  wall-less 
geometry. The end cells of each layer and the third layer are joined
together to form a veto layer for rejection of charged particles and 
background  produced by Compton scattering of gamma-rays.  The  odd 
and even cells of the first and second layers, which detect X-rays, 
are  connected together and operated in mutual  anticoincidence  to 
further  reduce the non-cosmic X-ray background. A sandwich  of  25 
micron  thick  aluminized  mylar  and  25  micron  thick   uncoated 
polypropylene serves as the X-ray entrance window. The  collimators 
with a field of view  of $2.3^\circ \times 2.3^\circ$ are made from
honeycomb shaped 
aluminum   coated  on both sides with a 6 micron  thick  layer  of 
silver.  The detectors and associated electronics are  so  designed 
that  each  PPC has a modular structure with  its  associated  high 
voltage  unit, signal processing electronics,  8086  microprocessor 
and a   memory of 4 MB. In the normal mode of operation count rates 
are  recorded from the first layer in 2-6 keV and 2-18  keV  bands, 
from  the second layer in  2-18 keV band and all count in the  veto 
layer  above  2  keV  with  an  integration  time  of  1  sec.  The 
integration  time can, however, be changed by command to  .01,  0.1 
or  10s.
The gas  gain  of each detector is measured  by  continuously 
monitoring  the  pulse  height  due  to  22.2  keV  X-rays  from  a 
collimated  Cadmium-109 radioactive source which irradiates only
the end cells of the veto layer. A more detailed description of
IXAE has been given by Agrawal et al. (\cite{agra:97}). 

     Since  the satellite is in a polar orbit and most of the
orbits  pass through the South Atlantic Anomaly (SAA)  region,  the 
useful data are obtained only when the satellite is usually in  the 
latitude range of 50$^\circ$ N to 30$^\circ$ S outside the SAA region.
The IRS-P3 is a three-axes  stabilized  satellite 
with  an  onboard star tracker which is used  for  acquisition  and 
pointing at a given X-ray source. 

\begin{figure}[t] 
\vskip 62mm
\includegraphics{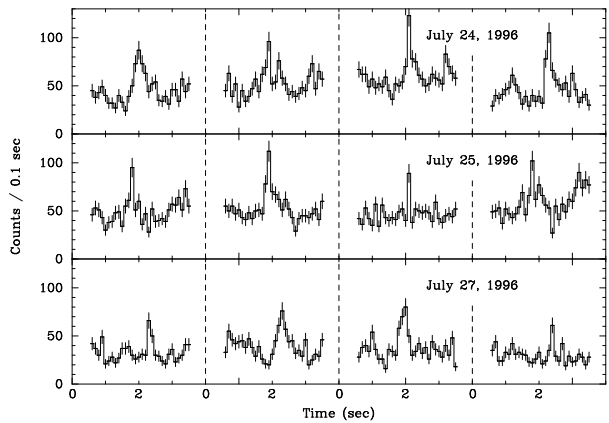} 
\caption{ The subsecond flux variations seen in GRS1915+105 with the Indian
X-ray Astronomy Experiment (IXAE). Observations were made with 100 msec time
resolution during 24-27 July 1996. Similar bursts were seen in all the detectors
and the combined data are plotted here. Each panel shows a few of the flares seen on 
the days mentioned in the figure. Data of 3 seconds are plotted with a bin size
of 100 msec around the flares. A factor of 2 or more increase in the X-ray flux
is seen for a duration of about 100 to 400 msec.}
\end{figure} 

     The X-ray instrument was first switched-on on  May 2, 1996 and 
its performance was verified by observing Cyg X-1 during May 2 -  9 
period.
The observed count rate for  Cyg  X-1  was 
about  500  counts  per  second indicating that it  was  in  a  low 
intensity  state. Subsequent pointing at Cyg X-1 during July  4-10, 
when  it  had made a transition to a bright state,  showed  a 
count  rate of about 1000 counts per sec. Preliminary results on
Cyg X-1 describing erratic variability on time scales of sub-second to second
are described by Agrawal et al. (\cite{agra:96a}). Observation of GRS 
1915+105  were  conducted 
during July 20-29, 1996. The target acquisition accuracy was 
better than 0.1$^\circ$. The background of the PPCs was measured by 
pointing  at  a  source-free region  near  GRS1915+105.  The  total 
background  count  rate  in 2-18 keV energy  band  from  the  three 
detectors is about 45 counts per second.

\begin{table}[h]
\caption{IXAE observations and results of GRS1915+105}
\begin{flushleft}
\begin{tabular}{llllllll}
\hline
Day of&From &To   &Useful&Cnts&QPO &rms\\
1996  &hh mm&hh mm&time  &per &freq& \%\\
July  &(UT) &(UT) &(sec) &sec &(Hz)&   \\
\hline
23      & 18 15 & 18 19 & ~240  &760    &       &       &\\
24      & 11 16 & 11 21 & ~120  &704    &0.68   & 8.29  &\\
24      & 12 43 & 13 02 & 1080  &664    &0.70   & 8.52  &\\
24      & 14 27 & 14 43 & ~900  &631    &0.70   & 9.13  &\\
25      & 12 33 & 12 39 & ~380  &755    &0.62   & 8.89  &\\
25      & 14 01 & 14 20 & 1080  &734    &0.62   & 9.21  &\\
25      & 15 44 & 16 02 & 1090  &779    &0.66   & 9.97  &\\
25      & 19 15 & 19 22 & ~420  &712    &0.70   & 9.44  &\\
25      & 20 59 & 21 12 & ~240  &707    &0.62   & 8.42  &\\
27      & 11 37 & 11 56 & 1080  &519    &0.74   & 9.00  &\\
27      & 13 21 & 13 37 & 1080  &467    &0.74   & 9.01  &\\
27      & 15 03 & 15 20 & 1140  &476    &0.82   & 9.43  &\\
\hline
\end{tabular}
\end{flushleft}
\end{table}

\section{Results }
 Even though GRS1915+105 was observed in  the  pointed-mode 
for  4 days during the period July 20-29, 1996 the useful data  are 
for about 8850 seconds due to various operational constraints. 
Almost all the count rates were acquired with an integration  time 
of  0.1 sec. The X-ray light curve for the entire period of  useful 
observations  showed  no large scale intensity variations  on  time 
scale  of a minute or longer. This is unlike the results   reported 
in  the higher energy band (\cite{cast:94}; \cite{sazo:94}). Analysis of data on shorter 
time  scale, however, shows pronounced variations on time scale of a second 
and  less. A few typical light curves showing  sub-second  intensity 
variations  are  shown  in  figure 1. These 
variations  were  detected  independently  in  each  detector  with 
similar count rate profiles. Further the veto layer count rates  do 
not  exhibit this kind of variability. It will be noticed from  the 
light  curves that GRS 1915+105 shows frequent flaring activity  on 
time  scales of less than a second and occasionally over 0.1  sec. 
During  the  flares the intensity varies by a factor of upto  3  in 
less than a second. From analysis of the flare frequency in GRS1915+105 
and  Cyg X-1, we find that flare occurrence in GRS1915+105  is  less 
common than that in Cyg X-1. Details of flaring activity in Cyg X-1 
will be reported in a separate paper (\cite{rao:97}). During the  sporadic 
intensity  variations, the light curves of the two sources  exhibit 
remarkable  similarity at time scale of 0.1 to 1 second. 
Power density spectrum  and  Fast  Fourier 
Transform (FFT) analysis showed no variability with a fixed  period 
indicating  that the source is not an X-ray pulsar. However, strong  Quasi-
Periodic  Oscillation (QPOs) were detected from GRS1915+105 in  all 
the data. The periodogram is shown in figure 2. The QPOs 
are  clearly  detected independently at the same frequency  in  the 
data  of each PPC as well as in the summed data. The QPO  frequency 
however  varies in an erratic manner from day to day. A summary  of 
the  observations and QPO characteristics is given in table I.
It will  be  noticed 
that  the  QPO frequency varied from 0.62 to 0.82 Hz.  The  rms
fraction in QPO is typically about 10\%.
Besides the 0.7 Hz peak the summed power spectrum
also shows another less prominent peak at $\approx$ 1.4 Hz, which can
be the first harmonic of the main peak.
The power spectrum is nearly flat between 0.02 to 0.5 Hz with a power law
index of -0.26 and becomes steep above the QPO frequency with a power law
index of -1.26 in the 2 to 5 Hz range.

\begin{figure}[t] 
\vskip 65mm
\includegraphics{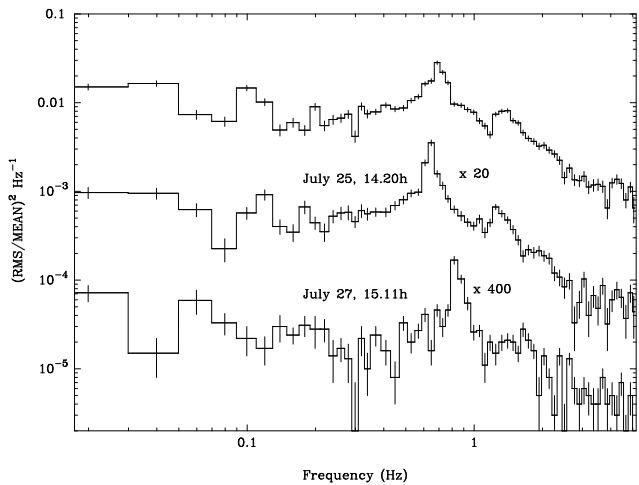} 
\caption{ The Power density spectrum of the X-ray source GRS1915+105 observed
with the Indian X-ray Astronomy Experiment (IXAE). The power density spectra 
of 130 independent data sets with 512 data points each, with a bin size of
100 msec were added to generate the plot at the top. A strong peak at a frequency
of 0.7 Hz is clearly visible. In the two plots below that two power
density spectra with peak at frequencies 0.62 Hz and 0.82 Hz respectively
show the frequency variation in the QPO peak. The times of observations are
given in the figure. The first harmonic at a frequency of twice that of the
main peak is seen in all the three plots}
\end{figure} 

Anticorrelation between the total intensity and hardness ratio (6-18 keV to
2-6 keV) is seen at very small time scale. Details of the analysis will be
reported in a separate paper.

\section{Discussion}

The  only reported  low   energy observations
(below 20 keV) are the one by Nagase et al. using the ASCA satellite (\cite{naga:94})
and by Greiner using the ROSAT satellite (\cite{grei:93}).
Recently, observations using PCA and ASM onboard the RXTE satellite have shown
intensity variation on a variety  of  time scales (\cite{grei:96}).
The maximum X-ray
luminosity for an assumed distance of 12.5 kpc (\cite{mira:94}),
is more than 10$^{39}$ erg s$^{-1}$ (\cite{sazo:94}). This
greatly exceeds the limiting Eddington luminosity for emission from
a neutron star surface of reasonable mass (\cite{lang:80}).
No coherent pulsations in the frequency range 0.001 to 0.1 Hz were
detected (\cite{fino:94}). Radio outbursts seen with the Green Bank
Interferometer (GBI) are found to be correlated with the X-ray flaring
seen by the BATSE onboard the CGRO (\cite{fost:96}) during a previous
outburst of the source.
The long time light curve of the source as seen by the All Sky Monitor (ASM)
on the RXTE after its recent outburst in January 1996, shows very strong
variability during May 15 to 30 June 1996 and August 15 to very recent time
and almost constant intensity between July 1 and August 15
1996 (http://space.mit.edu/XTE/ASM\_lc.html). Our observations were made
during the period of July 20-29 when ASM count rate shows no significant
intensity variations. Though the source was found to be bright in radio
during the current outburst too (\cite{fend:96}), there is no reported
radio observation during the rarely occuring constant intensity state
reported here.

	The subsecond time variability seen in GRS1915+105
indicates that the emission is from a compact region of size much smaller
than a light second. But the total X-ray intensity greatly exceeds the
Eddington luminosity from a neutron star with permitted mass limit.
Therefore the most likely place for the radiation to come from is the
accretion disk. In neutron star binaries, even a small magnetic field of
10$^{8}$ gauss is sufficient to keep the inner disk away from the neutron
star surface (\cite{fran:92}). Hence the subsecond variability that is
seen in GRS1915+105 indicates that the compact object in this system is
likely to be a black hole. The observed quasi-periodic oscillations when
compared to QPOs seen in other black hole X-ray binaries also support the
black hole picture.  

The stability of the QPO frequency over 4 days indicates that
the intensity oscillations are generated in an annular region in the disk.
If the QPOs were to arise due to blob of material orbiting in the disk, the
QPO frequency should have increased systematically with time
as the blob of material spirals towards the inner part of the disk.
If the inner disk is superheated it can emit very high energy
radiation upto gamma rays. Such an emission process is needed for the
plasmoids to be thrown away by the compact object in the form of jets by
radiation pressure along the axis of the system (\cite{lian:95}).  While calculating the
emission pattern from a disk around a black-hole two factors are to be
considered, the gravitational redshift experienced by the radiation from
the innermost disk and for super Eddington luminosity systems like
GRS1915+105 and the effect of radiation pressure in the inner disk structure.
If the radiation pressure is efficient in reducing the effective gravity
the radial structure of the disk can be very different from that of an ordinary
thin disk. These two factors, can lead to the fact that the efficient
radiation zone in a disk
around a black-hole can be somewhat further away from the event horizon.
The observed QPOs of 0.7 Hz and even smaller frequencies as seen by
PCA (http://heasarc.gsfc.nasa.gov/docs/xte/SOF/ toonews.html) can then be
from the most efficient zone of radiation in
the disk whose radius changes because of various disk instabilities.

In low mass X-ray binaries usually two types of QPOs are seen,
the nominal branch QPOs with a narrow frequency peak around 6 Hz with rms
variation of 1-3\% and the horizontal branch and flaring branch QPO peak
in the frequency range 15 - 50 Hz and somewhat larger rms variation.
Quasi-periodic oscillations are also seen in some of the pulsars but at
a lower frequency of 0.02 to 0.2 Hz and these can be explained using the beat
frequency model (\cite{klis:95}).
If the compact object in GRS1915+105
is a neutron star, the QPOs seen in this source are not like the ones in any
of the other neutron star sources. In many Black Hole Candidates (BHC)
QPOs associated with low frequency noise are seen at different frequencies
in the range of 0.04 to 6 Hz. The type of QPOs seen in GRS1915+105 are
similar to those seen in GX 339-4 (\cite{greb:91}).

Very strong subsecond intensity variations similar to GRS1915+105 are
also seen in other black-hole candidates like Cyg X-1 and GX 339-4 (\cite{klis:95}). Some
neutron star sources like Cir X-1 (\cite{toor:77}), 4U 1608-52 (\cite{klis:95}), V0332+53 (\cite{tana:83}) also have shown
subsecond variability but of smaller magnitude. So the short time variability seen
in the present observation alone does not prove that GRS1915+105 is a
black-hole source. However the Quasi-periodic oscillations at a frequency
of 0.7 Hz and its first harmonic brings out the remarkable similarity of
the power density spectrum (PSD) of this source with that of Cyg X-1 and
GS 1124-68 in their very high state (\cite{klis:95}).
The flatness of the PSD below the QPO peak and the steep
fall above the QPO frequency is also similar to that of other black-hole candidates in
their very high state. So the identical nature of the PSD of GRS 1915+105 to that
of Cyg X-1 and  GS 1124-68  and subsecond intensity variations by a factor of 2 or more makes
a strong case for this source to be a black hole. The hard X-ray tail
of the spectrum of GRS1915+105 also supports its black-hole nature. Simultaneous
observations in low and high energy X-rays in future will help in finding the
true nature of this source.
\begin{acknowledgements}
We gratefully acknowledge the strong support of Shri K. Thyagrajan, the Project
Director of IRS-P3 satellite and his entire team, Shri R. N. Tyagi,
Manager IRS programme, Shri R. Aravamudan, Director of ISAC for his support
as well as other technical
and engineering staff of ISRO Satellite Center in making the IXAE project a
success. We are in particular extremely thankful to the engineers, scientific and
technical staff of our group at TIFR, the group at Technical Physics Devision, ISAC
and ISRO tracking facility whose contributions were crucial to
the success of this experiment. 
\end{acknowledgements}

\end{document}